# Ion Induced Passivation of Grain Boundaries in Perovskite Solar Cells

Vikas Nandal and Pradeep R. Nair

Department of Electrical Engineering,

Indian Institute of Technology Bombay, Mumbai-400076, Maharashtra, India

*Abstract* – **Demonstration of high efficiency large area cells with excellent stability is an important requirement towards commercialization of perovskite solar cells (PSC). With reports of high quality perovskite grains, it is evident that the performance of such large area cells will be strongly influenced by phenomena like carrier recombination and ion migration at grain boundaries (GBs). Here, we develop a modeling framework to address performance limitation due to GBs in large area PSCs. Through detailed numerical simulations, we show that photo-carrier recombination has a non-trivial dependence on the orientation of GBs. Interestingly, we find that ions at GBs lead to significant performance recovery through field effect passivation, which is influenced by critical parameters like density and polarity of ions, and the location of GB. These results have interesting implications towards long term stability and hence are relevant for the performance optimization of large area polycrystalline based thin film solar cells such as PSCs, CIGS, CZTS, etc.**

Organic-inorganic hybrid perovskites are increasingly explored for various opto-electronic applications due to their excellent properties such as high absorption coefficient,[1] large diffusion length,[2–4] band gap tunability,[5–8] etc. Single junction perovskite solar cells (PSCs) already report efficiencies beyond 22 %.[9] Besides this, perovskite/silicon multi-junction solar cells have also achieved efficiencies of 25.2 % and 26 % with 2 T and 4 T configurations, respectively.[10,11] Low cost, ease of processing, and high flexibility are some of the other appealing attributes of this technology which makes perovskite an exciting and potential candidate for next-generation solar cells.[12,13]

In the development of PSCs, significant advances have been reported on the optimization of device architecture and fabrication processes.[13,14] Indeed, there have been reports of large grain sizes and excellent material purity as well.[15,16] While all these are encouraging, the quest for high efficiency large area perovskite solar cells are expected to be significantly affected by the presence of grain boundaries (GB, see Figure 1a) and the associated efficiency loss through carrier recombination,[17] ion migration,[18] etc. In this regard, we note that there have been reports of passivation of grain boundaries through organic materials[19–23] and techniques to enhance grain size[24–28] to suppress the effects of grain boundaries. Apart from these preliminary experiments, the device physics related to grain boundaries in perovskites and their impact on eventual performance is not very well explored in the literature. Therefore, it is imperative to investigate the role of grain boundaries on performance and hence identify schemes for further optimization.

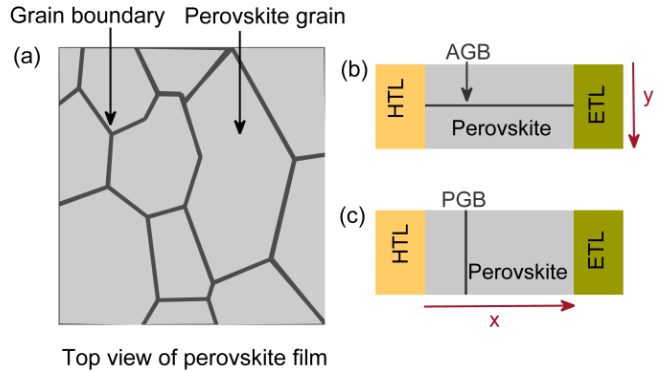

**Figure 1.** (a) Schematic of a typical top view of perovskite thin film illustrating different grains and grain boundaries. The device schematic of planar PIN based PSCs showing (b) aligned grain boundary AGB (i.e., GB parallel to current flow) and (c) perpendicular grain boundary PGB (i.e., GB perpendicular to current flow). The location of AGB and PGB is represented by $y$ and $x$, respectively.

In this letter, through detailed numerical simulations, we explore the effects of grain boundaries on the performance of planar PIN-based perovskite solar cells. A typical top-view of perovskite thin film is shown in Fig. 1a. With such a complex network of GBs, the problem requires carrier transport and optical simulations in 3 dimensions - which is indeed computationally very complex. To reduce the computational effort and to get better physical insights, here we map the complex network of grain boundaries to aligned grain boundary (AGB, see Fig. 1b) and perpendicular grain boundary (PGB, see Fig. 1c). Note that these assumptions on the orientation of GBs indeed allows us to obtain rich insights on the tradeoffs associated with the location and properties of GBs on device performance (other orientations could be similarly addressed, if need arises). Our detailed numerical simulations indicate that trapped ions inside GB can, in general, significantly mitigate the effects of trap assisted carrier recombination at GBs. Such ion induced passivation of GBs depends on additional parameters like ion polarity, type of GBs, material quality of perovskite grains and transport layers, etc. Below, we first describe the model system and the effect of GBs on the efficiency of perovskite solar cells.

We consider a planar PIN based perovskite solar cell as the model system where organic-inorganic $CH_3NH_3PbI_3$ perovskite, the active layer (band gap $E_g = 1.55\ eV$ corresponds to and thickness $l_{ac} = 300\ nm$), is sandwiched between carrier selective doped transport layers. Dark current is limited by charge carrier recombination in the active layer (AL) as is evident from energy level alignment of different layers (see Figure S1, in the supplementary material). Recombination of charge carriers inside AL is governed by trap-assisted (characterized by SRH carrier life time $\tau_B$), band to band radiative recombination ($B$ denotes the recombination parameter) and Auger recombination (with coefficient $A_n = A_p = A$).[29] We consider uniform optical generation rate $G$ across AL thickness and zero band offset at the interfaces of AL with carrier selective layers (i.e., ETL and HTL) for the extraction of photogenerated charge carriers by the respective contacts. The simulation methodology is well calibrated and reported in previous publications[30,31] and the simulation parameters are given in Table S1 (see supplementary information). Current voltage characteristics of the modeled device are obtained through self-consistent numerical solutions of drift-diffusion, continuity and Poisson's equations.

As mentioned before, grain boundaries and the associated effects like carrier recombination are of fundamental importance to perovskite based solar cells (and LEDs as well). Grain boundaries can, in general, result in (i) increased recombination of photogenerated carriers, (ii) can trap positive or negative ions (with ion density $N_I$), and (iii) facilitate migration of ions. The precise information on the material and electronic properties of grain boundaries such as effective thickness, carrier recombination velocities, band properties, density and nature of traps, etc. are indeed difficult to estimate and hence are also unavailable in literature. To make this problem conceptually and computationally tractable, here we define the grain boundary as a region of small thickness (2nm) with exactly the same properties as that of perovskite except for different trap assisted recombination rate (accounted through SRH carrier lifetime $\tau_{GB}$ inside GB) and the presence of ionic charge density $N_I$. Based on these assumptions, we now discuss the effect of AGB and PGB on perovskite solar cells.

**(i) Effects of Aligned Grain Boundary (AGB)**: Figure 1b shows the schematic of PSC with AGB (details given before). To study the effect of AGB on device performance, the carrier recombination rate inside GB is varied through the parameter $\tau_{GB}$ with fixed bulk carrier life time $\tau_B = 5\mu s$. In addition, we also vary the concentration of trapped ions by introducing fixed charge density ($N_I = \pm 10^{17}, \pm 10^{18}\ cm^{-3}$, here + and − sign indicate positive and negative ions, respectively) inside grain boundary region. Note that in the absence of any GBs, the model system is effectively 1-dimensional and the carrier densities are expected to vary only in the 'x' direction and not in the 'y' direction. The presence of AGBs will introduce variation in carrier density along the y-direction as well. Accordingly, the location of GBs could have some influence on the device characteristics. However, this spatial dependence is expected to be of the order of diffusion lengths and indeed we find that the characteristics are independent of the location of GBs once they are a few diffusion lengths away from the device boundaries.

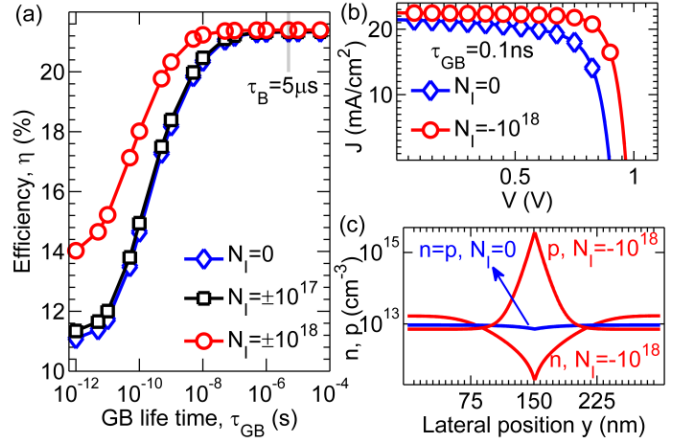

**Figure 2. Influence of AGB on efficiency of PSCs**. (a) Variation of device efficiency $\eta$ with SRH carrier life time $\tau_{GB}$ of AGB at different trapped ion density ($N_I = 0, \pm 10^{17}, \pm 10^{18}\ cm^{-3}$, here ± signs correspond to ion polarity) for fixed $\tau_B = 5\mu s$. The effect of trapped negative ions on (b) current voltage (JV) characteristics of PSC and (c) the carrier density (electron $n$, hole $p$) at $x \approx l_{ac}/2$ (the location of maximum recombination with $n = p$). The field effect due to negative ions passivate the AGB by reducing excess minority concentration (in this case electron density $n$). Device performance improves with the decrease in defect states (or increase in carrier life time $\tau_{GB}$) or increase in trapped ion density $N_I$ inside AGB.

Figure 2a shows the effect of $\tau_{GB}$ and $N_I$ (trapped ions) on device efficiency at a fixed location of AGB (at $y = 150\ nm$, where $y_{max} = 300\ nm$, is the width of the simulated device). Specifically, for $N_I = 0$, we find that the device efficiency improves with $\tau_{GB}$ and then saturates to the maximum achievable efficiency for a given $\tau_B$. Such performance improvement is due to decrease in recombination of photogenerated charge carriers inside AGB with the increase in $\tau_{GB}$. Interestingly, the presence of ions inside AGB (irrespective of the polarity) leads to an improvement in efficiency for all cases with $\tau_{GB} < \tau_B$. For example, for $\tau_{GB} = 0.1\ ns$, we find that the presence of ions improves $J_{sc}$, fill factor FF, and $V_{oc}$ (as shown in Fig. 2b). Field effect passivation of GB by ions (see Fig. 2c) results in the decrease of minority carrier concentration which in turn reduces charge carrier recombination inside AGB and hence increases the efficiency. We also observe that the magnitude of performance improvement due to trapped ions decreases with the increase in $\tau_{GB}$ (see Fig. 2a). Hence, device performance can be enhanced either by doing chemical passivation of AGB or by field passivation induced by trapped ions inside AGB. For instance, chemical treatment of AGB with organic materials can passivate the defect state density and improve the performance of the device[19–23]. Further, similar to CIGS treatment with alkali metals,[32–36] perovskite solar cells can be subjected to the doping of AGB with charged ionic species for the improvement in device performance.

**(ii) Effects of Perpendicular Grain Boundary (PGB):** The effects due to PGB can indeed be more complex than that of AGB as the carrier density varies significantly along the 'x' direction (see Fig. 4a). To illustrate this, here we consider different scenarios: Case A- electrostatics effect caused by trapped ions, Case B- effect due to increased charge carrier recombination in GBs, and Case C- Ion induced field effect passivation (i.e., combined effect of Case A and B). For Case A, we consider a fixed charge density of $N_I = +10^{18} cm^{-3}$ or $-10^{18} cm^{-3}$ with the carrier lifetime in GBs being the same as that of the perovskite bulk (i.e. $\tau_{GB} = \tau_B = 0.5\ \mu s$, i.e., GBs have the same recombination properties as the bulk of active layer). However, for Case B the carrier life time of PGB is less than perovskite bulk (i.e. $\tau_{GB} = 50\ ps$, $\tau_B = 5\mu s$) with the trapped ion density $N_I = 0$. For Case C, we consider the combined effect of increased carrier recombination and electrostatics of trapped ions ($\tau_{GB} = 50\ ps$, $\tau_B = 5\mu s$, and $N_I = \pm 10^{18} cm^{-3}$).

Figure 3(a, b) shows the effect of trapped ions, present at different location $x$ in perovskite layer, on the electrostatics of PSCs. The presence of ions inside PGB alters the electrostatics (hence the electric field profile) resulting in low field or diffused region and high field region in the perovskite layer as compared to $N_I = 0$. This modulation of the electric field inside perovskite layer depends on polarity and location of trapped ions. For example, the negative ions leads to low field or diffused region between GB and HTL/AL interface, whereas, the same ions results in high field region between GB and ETL/AL interface. The extent of low field or diffused region across active layer thickness increases as the location of PGB moves towards ETL (which is evident from Fig. 3a and b). Accordingly, the net carrier recombination can be affected depending on the relative magnitude of diffusion length ($\sqrt{D\tau}$)

vs. drift collection length ($\mu\tau E$) of photogenerated charge carriers ($D$ is diffusion constant, $\mu$ is carrier mobility, $\tau$ is effective carrier life time and $E$ is the electric field in perovskite layer). Therefore, the electrostatics of ions in GBs plays a crucial role in limiting the performance of PSCs. For instance, the increase in the extent of diffused region leads to increase in charge carrier recombination and hence the performance is expected to degrade with location $x$ of PGB with negative ions.

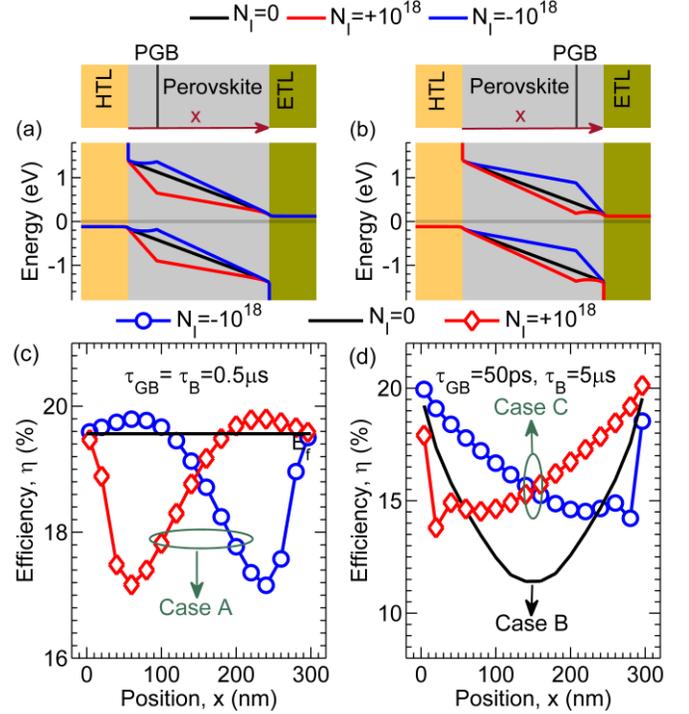

**Fig. 3. Effect of PGB on the performance of PSCs**. Part (a) and (b) illustrate the electrostatics effect of trapped ions (with density $N_I$ in $cm^{-3}$) present inside PGB at $x = 75\ nm$ and $x = 225\ nm$, respectively. Part (c) shows the electrostatics effect of trapped ions (Case A) on device efficiency. Part (d) shows the impact of carrier recombination (Case B – solid line) and ion induced passivation (Case C - symbols) at PGB on device performance. The device efficiency depends on carrier life time $\tau_{GB}$, location $x$ and polarity of ions. Field effect passivation of trapped ions lead to performance improvement or degradation of PSCs.

Figure 3c shows the electrostatics effect of trapped ions (i.e, Case A) on the performance of PSCs. Indeed we find that the device efficiency depends on the position $x$ along the AL thickness and polarity of ions. The efficiency is least affected due to the presence of ions at perovskite/contact layer interfaces for heavily doped contact layers. However, depending on polarity and position, the trapped ions present away from perovskite/contact layer interfaces can result in performance improvement or degradation. For instance, negative ions in the vicinity of HTL/AL interface show performance improvement while the same ions near AL/ETL interface results in performance degradation. In contrast, positive ions (shown in Figure 3a and 3b for different location $x$ of PGB) near HTL/AL or AL/ETL interface leads to performance loss or gain, respectively.

Figure 3d illustrates the effects of carrier recombination (Case B – solid line) and ion induced passivation (Case C - symbols) at GBs on the efficiency of PSCs. For Case B, we find that the performance reduces with increased carrier recombination ($\propto 1/\tau_{GB}$) inside PGB. Reduction in efficiency (due to degradation of $V_{oc}, J_{sc}$ and fill factor - see Figure S2 in the supplementary information) increases as PGB moves away from AL interfaces and reached a maximum for GBs at the middle of active layer. This trend could be anticipated from the nature of SRH trap assisted recombination and the carrier density profiles of PIN devices. Fig. 4a shows the carrier (electron $n$ and hole $p$) density profile inside perovskite layer under illumination condition with applied voltage $V = 0$. It is well known that the SRH recombination is at a maximum when n=p[29] and for a PIN structure with symmetric barriers the same happens at $x = l_{ac}/2$. Figure 4b shows the variation of net recombination rate (dominated by trap assisted SRH recombination rate) with AGB location. As expected the net recombination is at a maximum for $x = l_{ac}/2$ and this explains why PGBs near the middle of perovskite layer leads to maximum SRH recombination rate and hence the performance degradation (see case B, Fig. 3d).

Interestingly, we find that the presence of ions in PGBs result in significant performance recovery, as highlighted by Case C in Fig. 3d. The details of such performance recovery in terms of the carrier density profiles and recombination rates for a PGB at a specific location (at $x = 75$ nm) are provided in Fig. 4c. It is evident in Fig. 4c, that at the specific location of GB, the minority carrier concentration (in this case, the electrons) is reduced due to the ionic charge and hence this leads to a reduction in carrier recombination. Further, this reduction in carrier recombination is indeed bias dependent (see Fig. 4d), and at maximum power point condition (i.e., at V=0.9 V) the field effect passivation due to trapped ions lead to significant reduction in carrier recombination and hence a recovery in efficiency. We note that the performance improvement is strongly location dependent and in some cases could lead to degradation as well. For example, negative ions in GBs near the ETL/AL interface or positive ions in GBs near HTL/AL interface lead to performance degradation (see Fig. 3d). In all other cases, ions in GBs lead to performance improvement. In general, it is advisable to have negative ions in GBs towards the HTL/AL interface and positive ions in GBs near the ETL/AL interface to maximize the performance recovery from the ill effects of increased carrier recombination at GBs. Additional discussion on the effect of material properties of perovskite and contact layers on the influence of ions in GBs (Case A) is provided in Figure S3 of supplementary materials.

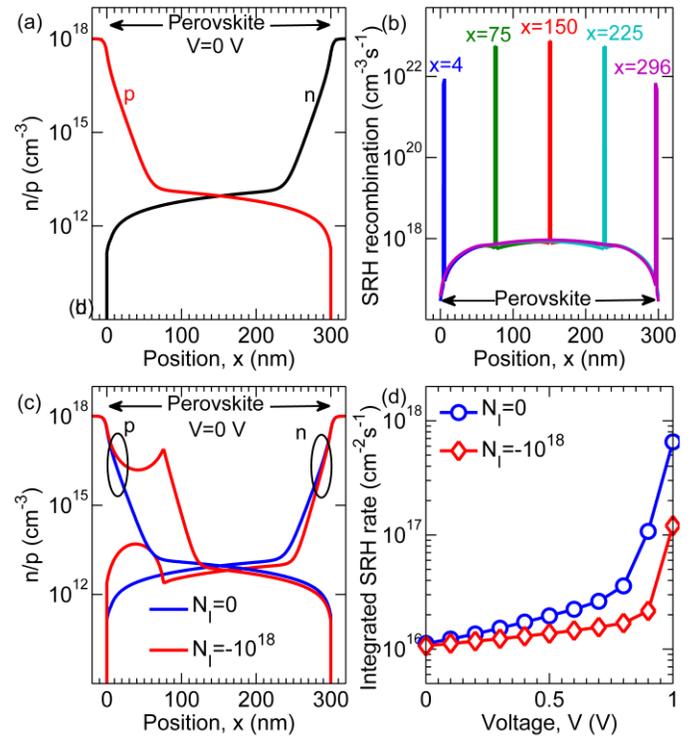

**Fig. 4. Ion-induced field effect passivation of the perpendicular grain boundary of PSCs.** (a) photogenerated charge carrier density profiles for $V = 0$ in the absence of GBs. (b) SRH recombination rate inside perovskite layer for different location $x$ (4, 75, 150, 225, 296 nm) of PGBs. Here, the carrier life time of bulk and PGB is $5\mu s$ and $50\ ps$, respectively. Maximum carrier recombination occurs when PGB is located in the middle of perovskite layer. (c) The effect of negative ions (with density $N_I$ and location $x = 75$ nm) on charge carrier density inside perovskite layer. Part (d) illustrates the bias dependent net carrier recombination in the GBs. The presence of negative ions decreases the electron density inside GB which leads to a reduction in net SRH recombination rate and hence improves the device performance.

Our results indicate that grain boundaries play a crucial role in limiting the performance of PSCs. Indeed, the charge carrier recombination through defect states of grain boundaries leads to performance loss of PSCs. Such detrimental effect on performance can be suppressed by chemical or electrical passivation of GBs. Various organic materials can be used to passivate the defect states to increase the carrier lifetime. Similar to alkali treatment of CIGS-based solar cells,[32–36] the grain boundary can also be filled with immobile ions to reduce minority carrier concentration. These treatments of perovskite material will reduce trap/defect-assisted SRH recombination of charge carrier inside GBs, and hence, improve the performance of PSCs. Our results can be beneficial towards the development of large area PSCs where the number of grain boundaries can be substantial which might limit the device performance. Further, the characterization schemes such as electroluminescence and photoluminescence can be used to identify and map out the location of grain boundary inside active layer and any associated passivation effects.[37,38] Similar to cracks identification in Silicon, the presence of defect states in GBs leads to low luminescence intensity as compared to perovskite bulk. In addition, our results can be used as a theoretical platform to understand effects like grain boundary induced hysteresis and long-term reliability or stability in PSCs. It is reported in the literature that ion migration could be significant through GBs.[18,39–42] Therefore, the dynamics of ion trapping or release from GBs can be explored to identify the physical mechanism behind such effects.

In conclusion, we study the role of grain boundaries (i.e. aligned and perpendicular) on the performance of PSCs. Our results indicate that GBs degrades the device performance and the magnitude of performance loss depends on carrier recombination rate in the GB along with orientation and location of GB. However, the trapped ions inside GB induce partial or full recovery of device performance through field effect passivation. The magnitude of such recovery depends on orientation, location, ion density, and polarity. Besides gain in performance, the trapped ions inside GB can indeed lead to long-term degradation of PSCs, which could be of further interest.

ACKNOWLEDGEMENT

This paper is based upon work supported under the US-India Partnership to Advance Clean Energy-Research (PACE-R) for the Solar Energy Research Institute for India and the United States (SERIIUS), funded jointly by the U.S. Department of Energy (Office of Science, Office of Basic Energy Sciences, and Energy Efficiency and Renewable Energy, Solar Energy Technology Program, under Subcontract DE-AC36-08GO28308 to the National Renewable Energy Laboratory, Golden, Colorado) and the Government of India, through the Department of Science and Technology under Subcontract IUSSTF/JCERDC-SERIIUS/2012 dated 22nd November 2012. The authors also acknowledges Center of Excellence in Nanoelectronics (CEN) and National Center for Photovoltaic Research and Education (NCPRE), IIT Bombay for computational facilities.

REFRENCES

[1] S. De Wolf, J. Holovsky, S.-J. Moon, P. Löper, B. Niesen, M. Ledinsky, F.-J. Haug, J.-H. Yum, and C. Ballif, J. Phys. Chem. Lett. **5**, 1035 (2014).

[2] S. D. Stranks, G.E. Eperon, G. Grancini, C. Menelaou, M.J.P. Alcocer, T. Leijtens, L.M. Herz, A. Petrozza, and H.J. Snaith, Science **342**, 341 (2013).

[3] G. Xing, N. Mathews, S. Sun, S.S. Lim, Y.M. Lam, M. Grätzel, S. Mhaisalkar, and T.C. Sum, Science **342**, 344 (2013).

[4] E. Edri, S. Kirmayer, A. Henning, S. Mukhopadhyay, K. Gartsman, Y. Rosenwaks, G. Hodes, and D. Cahen, Nano Lett. **14**, 1000 (2014).

[5] M.R. Filip, G.E. Eperon, H.J. Snaith, and F. Giustino, Nat. Commun. **5**, 5757 (2014).

[6] N.K. Kumawat, A. Dey, A. Kumar, S.P. Gopinathan, K.L. Narasimhan, and D. Kabra, ACS Appl. Mater. Interfaces **7**, 13119 (2015).

[7] N.K. Kumawat, A. Dey, K.L. Narasimhan, and D. Kabra, ACS Photonics **2**, 349 (2015).

[8] S.A. Kulkarni, T. Baikie, P.P. Boix, N. Yantara, N. Mathews, and S. Mhaisalkar, J. Mater. Chem. A **2**, 9221 (2014).

[9] W.S. Yang, B.-W. Park, E.H. Jung, N.J. Jeon, Y.C. Kim, D.U. Lee, S.S. Shin, J. Seo, E.K. Kim, J.H. Noh, and S. Il Seok, Science **356**, 1376 (2017).

[10] F. Sahli, J. Werner, B.A. Kamino, M. Bräuninger, R. Monnard, B. Paviet-Salomon, L. Barraud, L. Ding, J.J. Diaz Leon, D. Sacchetto, G. Cattaneo, M. Despeisse, M. Boccard, S. Nicolay, Q. Jeangros, B. Niesen, and C. Ballif, Nat. Mater. (2018).

[11] C.O. Ramirez Quiroz, Y. Shen, M. Salvador, K. Forberich, N. Schrenker, G.D. Spyropoulos, T. Heumuller, B. Wilkinson, T. Kirchartz, E. Spiecker, P.J. Verlinden, X. Zhang, M.A. Green, A. Ho-Baillie, and C.J. Brabec, J. Mater. Chem. A **6**, 3583 (2018).

[12] J. You, Z. Hong, Y.M. Yang, Q. Chen, M. Cai, T. Song, and C. Chen, ACS Nano **8**, 1674 (2014).

[13] H.J. Snaith, J. Phys. Chem. Lett. **4**, 3623 (2013).

[14] M. a. Green, A. Ho-Baillie, and H.J. Snaith, Nat. Photonics **8**, 506 (2014).

[15] W. Nie, H. Tsai, R. Asadpour, J.-C. Blancon, A.J. Neukirch, G. Gupta, J.J. Crochet, M. Chhowalla, S. Tretiak, M.A. Alam, H.-L. Wang, and A.D. Mohite, Science **347**, 522 (2015).

[16] S.D. Stranks and H.J. Snaith, Nat. Nanotechnol. **10**, 391 (2015).

[17] C. Feilong, Y. Yu, Y. Jiaxu, W. Pang, W. Hui, G.R. S., L. Dan, and W. Tao, Adv. Funct. Mater. **0**, 1801985 (2018).

[18] Y. Shao, Y. Fang, T. Li, Q. Wang, Q. Dong, Y. Deng, Y. Yuan, H. Wei, M. Wang, A. Gruverman, J. Shield, and J. Huang, Energy Environ. Sci. **9**, 1752 (2016).

[19] D.S. Lee, J.S. Yun, J. Kim, A.M. Soufiani, S. Chen, Y. Cho, X. Deng, J. Seidel, S. Lim, S. Huang, and A.W.Y. Ho-Baillie, ACS Energy Lett. **3**, 647 (2018).

[20] Y. Shao, Z. Xiao, C. Bi, Y. Yuan, and J. Huang, Nat. Commun. **5**, 5784 (2014).

[21] N. Tianqi, L. Jing, M. Rahim, L. Jianbo, B. Dounya, Z. Xu, H. Hanlin, Y. Zhou, A. Aram, Z. Kui, and L.S. (Frank), Adv. Mater. **30**, 1706576 (2018).

[22] J. Xu, A. Buin, A.H. Ip, W. Li, O. Voznyy, R. Comin, M. Yuan, S. Jeon, Z. Ning, J.J. McDowell, P. Kanjanaboos, J.-P. Sun, X. Lan, L.N. Quan, D.H. Kim, I.G. Hill, P. Maksymovych, and E.H. Sargent, Nat. Commun. **6**, 7081 (2015).

[23] B. Yang, O. Dyck, J. Poplawsky, J. Keum, A. Puretzky, S. Das, I. Ivanov, C. Rouleau, G. Duscher, D. Geohegan, and K. Xiao, J. Am. Chem. Soc. **137**, 9210 (2015).

[24] H. Zhou, Q. Chen, G. Li, S. Luo, T. Song, H.-S. Duan, Z. Hong, J. You, Y. Liu, and Y. Yang, Science **345**, 542 (2014).

[25] C. Bi, Q. Wang, Y. Shao, Y. Yuan, Z. Xiao, and J. Huang, Nat. Commun. **6**, 7747 (2015).

[26] X. Zhengguo, D. Qingfeng, B. Cheng, S. Yuchuan, Y. Yongbo, and H. Jinsong, Adv. Mater. **26**, 6503 (2014).

[27] D. Liu, L. Wu, C. Li, S. Ren, J. Zhang, W. Li, and L. Feng, ACS Appl. Mater. Interfaces **7**, 16330 (2015).

[28] Q. Dong, Y. Yuan, Y. Shao, Y. Fang, Q. Wang, and J. Huang, Energy Environ. Sci. **8**, 2464 (2015).

[29] S.M. Sze, *Physics of Semiconductor Devices* (John Wiley & Sons, 1981).

[30] S. Agarwal and P.R. Nair, Appl. Phys. Lett. **107**, 123901 (2015).

[31] V. Nandal and P.R. Nair, ACS Nano **11**, 11505 (2017).

[32] N. Nicoara, T. Lepetit, L. Arzel, S. Harel, N. Barreau, and S.

Sadewasser, Sci. Rep. **7**, 41361 (2017).

[33] P. Reinhard, B. Bissig, F. Pianezzi, E. Avancini, H. Hagendorfer, D. Keller, P. Fuchs, M. Döbeli, C. Vigo, P. Crivelli, S. Nishiwaki, S. Buecheler, and A.N. Tiwari, Chem. Mater. **27**, 5755 (2015).

[34] F. Pianezzi, P. Reinhard, A. Chirila, B. Bissig, S. Nishiwaki, S. Buecheler, and A.N. Tiwari, Phys. Chem. Chem. Phys. **16**, 8843 (2014).

[35] D. Rudmann, A.F. da Cunha, M. Kaelin, F. Kurdesau, H. Zogg, A.N. Tiwari, and G. Bilger, Appl. Phys. Lett. **84**, 1129 (2004).

[36] L. Anke, W. Roland, and P. Michael, Phys. Status Solidi – Rapid Res. Lett. **7**, 631 (2013).

[37] T. Fuyuki, H. Kondo, T. Yamazaki, Y. Takahashi, and Y. Uraoka, Appl. Phys. Lett. **86**, 262108 (2005).

[38] T. Trupke, R.A. Bardos, M.C. Schubert, and W. Warta, Appl. Phys. Lett. **89**, 44107 (2006).

[39] H. Mehrer, *Diffusion in Solids: Fundamentals, Methods, Materials, Diffusion-Controlled Processes* (Springer Berlin Heidelberg, 2007).

[40] Y. Bin, B.C. C., H. Jingsong, C. Liam, S. Xiahan, U.R. R., J. Stephen, K.S. V., B. Alex, J. Jacek, G.D. B., S.B. G., X. Kai, and O.O. S., Adv. Funct. Mater. **27**, 1700749 (2017).

[41] Y.J. S., S. Jan, K. Jincheol, S.A. Mahboubi, H. Shujuan, L. Jonathan, J.N. Joong, S.S. Il, G.M. A., and H.-B. Anita, Adv. Energy Mater. **6**, 1600330 (2016).

[42] Y. Yuan and J. Huang, Acc. Chem. Res. **49**, 286 (2016).